\documentclass{aa} 
\usepackage{graphicx}
\usepackage{txfonts}
\usepackage{natbib}
\bibpunct{(}{)}{;}{a}{}{,} 
\usepackage{color}
\usepackage{paralist}
\usepackage{verbatim}
\usepackage{ulem}

\newcommand{\mch}{\ensuremath{\mathrm{M}_\mathrm{Ch}}}
\newcommand{\msun}{\ensuremath{\mathrm{M}_\odot}}
\newcommand{\nuc}[2]{\ensuremath{^{#1}\mathrm{#2}}}
\newcommand{\kms}{\ensuremath{\mathrm{km}\,\mathrm{s}^{-1}}}

\newcommand{\gcc}{\ensuremath{\mathrm{g\ cm^{-3}}}}
\newcommand{\K}{\ensuremath{\mathrm{K}}}
\newcommand{\code}[1]{\textsc{#1}}
\newcommand{\band}[1]{\textit{#1}}
\newcommand{\foe}{\ensuremath{10^{51}~\mathrm{erg}}}

\graphicspath{{plots/}}

\begin{document}
\title{Type Ia supernovae from exploding oxygen-neon white dwarfs}


\author{Kai S. Marquardt$^{1,2}$, Stuart A. Sim$^{3,4}$, Ashley J. Ruiter$^{5,4}$, Ivo R. Seitenzahl$^{5,4,1}$, 
Sebastian T. Ohlmann$^{1,2}$, Markus Kromer${^6}$, R{\"u}diger Pakmor$^{2}$, Friedrich K. R{\"o}pke$^{7,2}$}

\institute{
  Institut f\"ur Theoretische Physik und Astrophysik,
  Universit\"at W\"urzburg, Campus Hubland Nord,
  Emil-Fischer-Str. 31,\\
  D-97074 W\"urzburg, Germany\\
  \email{kmarquardt@astro.uni-wuerzburg.de}
  \and
  Heidelberger Institut f{\"u}r Theoretische Studien, Schloss-Wolfsbrunnenweg 35, D-69118 Heidelberg, Germany
  \and
  School of Mathematics and Physics,
  Queen's University Belfast University Road Belfast, 
  Northern Ireland BT7 1NN United Kingdom
  \and 
  ARC Centre of Excellence for All-Sky Astrophysics (CAASTRO)
  \and
  Research School of Astronomy and Astrophysics, Mount Stromlo 
  Observatory, Weston Creek, ACT 2611, Australia
  \and
  The Oskar Klein Centre \& Department of Astronomy,
  Stockholm University, AlbaNova, SE-106 91 Stockholm, Sweden
  \and
  Zentrum f\"ur Astronomie der Universit\"at Heidelberg,
  Institut f\"ur Theoretische Astrophysik, Philosophenweg 12, D-69120 Heidelberg, Germany
}

\date{Received xxxx xx, xxxx / accepted xxxx xx, xxxx}


\abstract
{
  The progenitor problem of Type Ia supernovae (SNe~Ia) is still unsolved.
  Most of these events are thought to be explosions of carbon-oxygen (CO)
  white dwarfs (WDs), but for many of the explosion scenarios,
  particularly those involving the externally triggered detonation of a sub-Chandrasekhar mass WD (sub-\mch~WD),
  there is also a possibility of having an oxygen-neon (ONe) WD as progenitor.
}
{
  We simulate detonations of ONe WDs and calculate synthetic
  observables from these models.
  The results are compared with detonations in CO WDs of
  similar mass and observational data of SNe Ia.
}
{
  We perform hydrodynamic
  explosion simulations of detonations in initially hydrostatic ONe WDs for a range of masses below
  the Chandrasekhar mass (\mch), followed by detailed nucleosynthetic postprocessing with a 384-isotope
  nuclear reaction network. The results are used to calculate
  synthetic spectra and light curves, 
  which are then compared with observations of SNe~Ia.
  We also perform binary evolution calculations
  to determine the number of SNe~Ia involving ONe WDs relative to the number of other promising progenitor channels.
}
{
  The ejecta structures of our simulated detonations in sub-\mch~ONe
  WDs are similar to those from CO WDs. There
  are, however, small systematic deviations in the mass fractions and
  the ejecta velocities.
  These lead to spectral features that are systematically less blueshifted.
  Nevertheless, the synthetic observables of
  our ONe WD explosions are similar to those obtained from
  CO models.
}
{
  Our binary evolution calculations show that a significant fraction (3-10\%)
  of potential progenitor systems should contain an ONe WD.
  The comparison of our ONe models with our CO models of comparable mass (${\sim}1.2\,\msun$) shows that the less blueshifted spectral features fit  the observations better, although they are too bright for normal SNe~Ia.
}
\keywords{supernovae: general -- nuclear reactions, nucleosynthesis,
  abundances -- hydrodynamics, radiative transfer -- white dwarfs -- Stars: evolution }

\titlerunning{ONe White Dwarfs as SN Ia progenitors}
\authorrunning{Marquardt et al. 2015} 
\maketitle
%

\section{Introduction}
Type Ia supernovae (SNe~Ia) are believed to result from thermonuclear explosions of white dwarf
(WD) stars \citep{hoyle1960a} in binary systems. The parameters of the
progenitor systems \citep{wang2012b}, however, and  the details of the
explosion mechanism \citep[e.g.][]{hillebrandt2000a} remain unclear.
Several scenarios 
hold promise for explaining
normal or peculiar SNe~Ia 
\citep[see e.g.][]{hillebrandt2013a}.
Typically, the exploding WD in all these scenarios is assumed to be
a carbon-oxygen (CO) WD.
Here, we explore the possibility of explosions in ONe WDs.

For a long time, explosions of  near-Chandrasekhar mass WDs (near-\mch~WDs) formed in
the single-degenerate progenitor channel were
the favoured model of SNe~Ia \citep[see][]{hillebrandt2000a}.
In this scenario, ONe WDs are excluded because, although electron captures on \nuc{24}{Mg} and \nuc{20}{Ne} can ignite a nuclear flame, further electron captures in the O-burning ashes lead to loss of pressure support and collapse \citep{miyaji1980a,miyaji1987a}.
They are therefore expected to form neutron stars rather than being disrupted in a
thermonuclear explosion as their mass approaches \mch\ \citep{nomoto1984b,nomoto1987a,nomoto1991a}.
However, there is mounting evidence for other progenitor
channels contributing to (or dominating) the sample of SNe~Ia
\citep[e.g.][]{stritzinger2006a,ruiter2009a,
  gilfanov2010a,sim2010a,scalzo2014a,scalzo2014b}. Many of these alternatives involve
detonations in sub-\mch~WDs. This again raises the
question of whether ONe
WDs contribute to the progenitor population,
since sub-\mch~configurations
are stable against gravitational collapse and 
detonations propagate rapidly enough such that
electron captures do not lead to a collapse.
This, however, requires the triggering of a detonation in ONe WD
matter, which may be possible but has not yet been proven to work
\citep{shen2014a}. Detonations in ONe WDs could be
ignited in the double detonation \citep{livne1990a,livne1990b,livne1991a,woosley1994b,fink2007a,
    fink2010a, woosley2011b,moll2013a} or violent merger
  \citep{pakmor2010a, pakmor2012a, pakmor2013a, moll2013b} scenarios.
Despite the uncertainties related to their ignition,
we here investigate the question of how, if they do occur,
such events might differ from those with CO WD progenitors and whether
they might be identifiable as a subpopulation of SNe~Ia.

\section{Population synthesis and the origin of ONe WDs}
\label{sec:population_synthethis_and_the_origin_of_one_wds}
\citet{garcia-berro1997a} and \citet{gil-pons2001a} investigated
asymptotic giant branch (AGB) stars
with zero-age main sequence (ZAMS) masses of $9$--$10\,\msun$;
the first study is for a single star while the second paper describes the evolution of a close binary system.
These stars produce ONe WDs with total masses of about $1.1\,\msun$.  
They are mainly composed of \nuc{16}{O} and \nuc{20}{Ne},
but they also contain some \nuc{12}{C}, the exact amount of which
depends on the initial model.\footnote{In addition to effects discussed here, another factor that influences the final composition of the AGB core is the carbon burning rate \citep{chen2014a}.} 
For a $9\,\msun$ ZAMS progenitor, for instance, the \nuc{12}{C}
mass fraction in the WD material can be up to ${\sim}0.05$, but it varies with
radius. This C admixture is important for the initiation and subsequent propagation of the detonation because it acts as an accelerant for neon burning.
However, successful initiation and propagation of the detonation are assumptions in the present work. The viability of these can only be addressed by spatially
resolved direct numerical simulations of the hydrodynamics coupled to a full nuclear reaction network, unfortunately still out of reach for full star explosion simulations.

\begin{figure}
  \begin{center}
     \includegraphics[width=\linewidth]{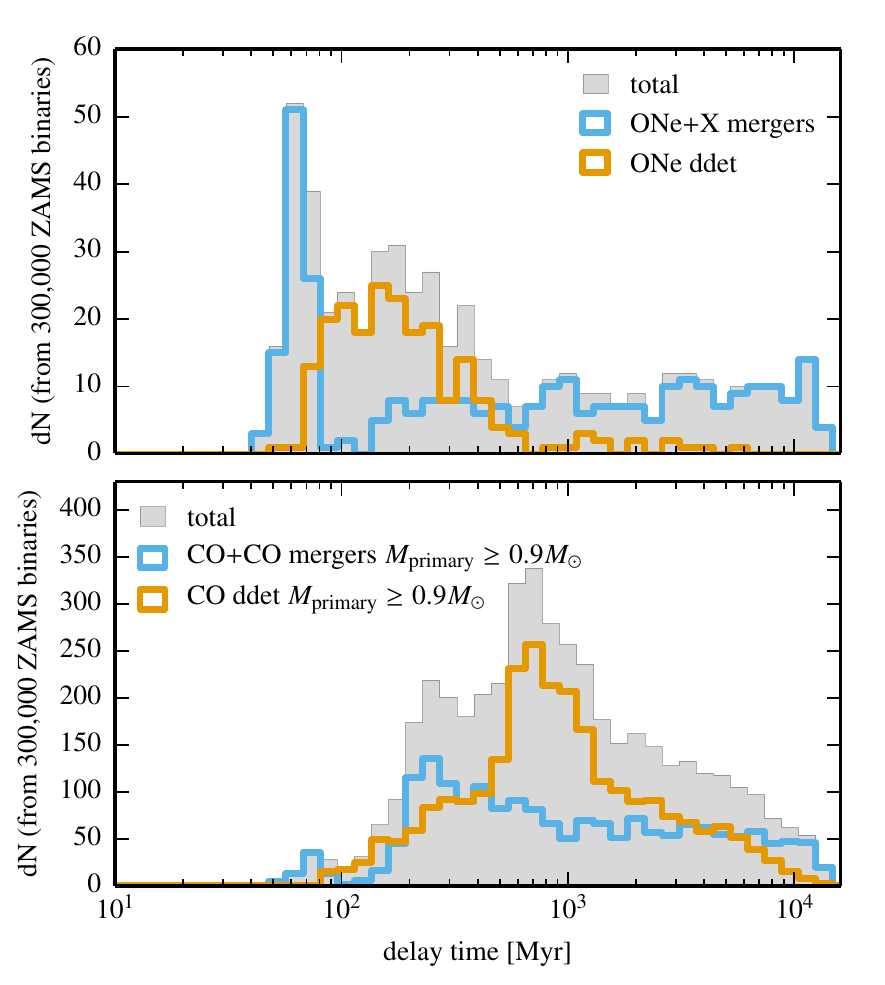}
     \caption{
       Delay time distributions for a subset of SN~Ia progenitors shown in Table~\ref{tab:rates}. Top panel: mergers that involve one or more ONe WD (blue), and double detonations where the accretor is an ONe WD (orange). Bottom panel: mergers that involve two CO WDs where the primary (more massive) WD is $\ge 0.9$ \msun\ (blue), and double detonations where the accretor is a CO WD with mass $\ge 0.9$ \msun\ (orange). For both panels the combined DTD is shown in grey. 
            }
       \label{fig:dtd}
  \end{center}
\vspace{-0.0cm}
\end{figure}

The WDs formed in binary systems cover a wider range of masses due to mass gain/loss
from/to the binary companion during stellar evolution. Depending on previous mass transfer
episodes, the ONe WD can easily have a mass ranging from $1.08\,\msun$ up to \mch,
while CO WDs can be formed with masses as high as ${\sim} 1.25\,\msun$ \citep[see][]{hurley2000a}.
We use the binary population synthesis code \code{StarTrack} \citep{belczynski2002a,belczynski2008a} to predict the number of
potential SN~Ia progenitors that involve ONe WDs both for dynamical (mergers)
and non-dynamical (classic double detonation) scenarios\footnote{In the case of classic double detonations,
  the mass ratio is often sufficiently far from unity to enable stable mass transfer to proceed once the larger WD fills its Roche lobe.}, and show their delay time distribution (DTD; Figure~\ref{fig:dtd}).
The results are summarized in Table~\ref{tab:rates}. 
We find that averaged over a Hubble time, the total rate of CO$+$CO mergers
is $1.06 \times 10^{-13}$ \msun$^{-1}$ yr$^{-1}$, where the mass represents mass born in stars.
This value is very close to the estimated SN~Ia rate in
Milky Way-like galaxies \citep[$1.1 \times 10^{-13}$ \msun$^{-1}$ yr$^{-1}$, see][]{badenes2012a}.
Thus, we use the total number of all CO$+$CO mergers as a reference point when comparing total numbers for various progenitor scenarios in Table~\ref{tab:rates}.

\begin{table}[h]
  \caption {Relative rates averaged over a Hubble time for theoretically predicted explosion scenarios that
      may lead to SNe~Ia. Data are extracted from the P-MDS model of
      \citet{ruiter2014a}. 
      All values have been normalized to the number of carbon-oxygen WD
      mergers (full mass range).
      The double-detonation systems are denoted by ddet.
      The accretion induced collapse systems arise from ONe WDs that collapse
      to a neutron star as they approach \mch\ while
      accreting stably from a stellar companion. Entries marked with an asterisk are the systems for which we show delay times in Figure~\ref{fig:dtd}.}
 \begin{center}
  \begin{tabular}{lc}
    \hline\hline
    SN~Ia progenitor scenario & rel. rate \\
    \hline
    CO$+$CO mergers (all) & 1.0\\
    CO$+$CO mergers (primary mass $>0.9\,\msun$)* & 0.27\\
    ONe$+$X mergers (all)* & 0.04\\
    ddet (CO primary, all) & 0.86\\
    ddet (CO primary mass $> 0.9\,\msun$)*  & 0.37 \\
    ddet (ONe primary; all)* &  0.03\\
    Chandrasekhar-mass CO WD (SD) & 0.01\\
    accretion-induced collapse in ONe WD (AIC) & 0.02 \\
    \hline
  \end{tabular}
  \end{center}
    \label{tab:rates}
\end{table}

If we compare the systems of ONe+ONe WDs combined with the CO$+$ONe
WDs (denoted  ONe$+$X mergers) that end up
in a merger scenario, their fraction is 4\% of the CO$+$CO mergers.
Of course, not all CO$+$CO WD mergers involve primaries massive enough to
lead to a bright SN~Ia in the violent merger scenario \citep{pakmor2010a,ruiter2013a}. If we restrict
the sample of CO$+$CO mergers to those where the primary WD has at least
$0.9\,\msun$, the ONe$+$X merger fraction is as high as 16\%.
We find that the number of binary systems involving ONe WDs that may lead to
classical (non-dynamical mass transfer) double-detonations are about 8\% of
the classical CO double-detonations
\citep[see][]{ruiter2014a}.
Here, for CO WD primaries, we assume
that only the systems in which the primaries are more massive than $0.9\,\msun$ will
potentially lead to thermonuclear events that are bright enough to be
considered SNe~Ia \citep{sim2010a,ruiter2014a}. The low-mass systems are
  thus excluded from being considered likely SN~Ia candidates; however, we include
  their numbers in  Table~\ref{tab:rates} for completeness.
For the systems involving ONe primaries, the lowest mass primary is
${\sim}1.1\,\msun$, thus we include all of them.
We note that for the population synthesis model presented, the number of
ONe double-detonations is a factor of three higher than 
single degenerate (SD) systems involving a \mch~CO WD.
Taken together, these relative rate estimates suggest that potential progenitor systems
in which an ONe WD explodes are frequent enough to constitute a
substantial subset of thermonuclear explosions.

In Figure~\ref{fig:dtd}, top panel, we show the DTDs for double detonations in ONe WDs and double WD mergers involving one or more ONe WD. The majority of ONe double detonations (orange) have delay times $< 650$ Myr. The donor stars in these binaries are naked helium-burning stars that were formed during the second common envelope phase that is encountered during the evolution. A small number of systems with delay times $\gtrsim 1$ Gyr involve helium WD donors. Most of the mergers (blue) are between ONe and CO WDs, though in some cases the mergers are ONe+ONe (14\%). Taking the mergers alone, the DTD shape does not resemble a power law, which has been extensively accredited to merging CO WDs in the literature \citep[see discussion in][]{totani2008a}.  In the bottom panel of Figure~\ref{fig:dtd} we show the DTD for binaries involving CO WDs where the primary CO WD mass for both double detonations and mergers is $\ge 0.9$ \msun. The double detonations show a very different DTD shape than that of the ONe systems; the peak at ${\sim} 700$ Myr is simply not found in systems with ONe primaries (instead there is a peak ${\sim} 200$ Myr). The lack of a later DTD peak for double detonations involving ONe WDs is due to the fact that events with delay times ${\gtrsim} 800$ Myr typically involve degenerate (helium WD) donors \citep{ruiter2014a}, which are not produced as frequently in binaries involving the more massive ONe primaries. For mergers involving two CO WDs, again, the DTD shape is quite different in comparison to mergers involving an ONe WD. It is worth noting that the number of ultra-prompt ($< 100$ Myr) mergers in binaries involving an ONe WD is higher than that of CO+CO systems, even though the latter (even with the adopted mass cut) outnumber the ONe+X mergers by a factor of ${\sim}7$. These ultra-prompt events originate from progenitors that undergo two common envelope events where the same star loses its envelope twice \citep[see][for discussion]{ruiter2013a}. 
  Unlike the mergers with ONe WDs, the CO+CO merger DTD peaks around $200-300$ Myr, consistent with the previous results of \citet{ruiter2013a}. However, cutting out the lower mass CO WD primaries leads to a slightly flatter DTD shape than found when including a larger binary sample, where a $t^{-1}$ power law is typically found.
  
  We do not include mergers between CO WDs and He-rich WDs, which may also lead to double detonations. We find that such systems always have CO WD masses below 0.7 \msun\ at the time of the merger. Typical masses for systems that lead to unstable mass transfer and merge are ${\lesssim} 0.35$ \msun\ and ${\sim} 0.6$ \msun\ for He and CO WDs, respectively. Whether or not a detonation of the CO core would be triggered by a He-detonation in such low-mass systems is still uncertain \citep{sim2012a,shen2014a}, and in fact some of these mergers may lead to the formation of RCrB stars rather than to thermonuclear explosions \citep{webbink1984a}. Either way, compared to SNe~Ia, successful CO ignition in such systems would produce fainter and faster evolving thermonuclear transients \citep{sim2012a} and would therefore not contribute to the SN~Ia rate.

\section{Explosions of ONe white dwarfs}
\label{sec:explosion}

\subsection{Numerical methods}
\label{subsec:numerical_methods}
For our explosion simulations we use the Eulerian hydrodynamics code \code{leafs} \citep{reinecke1999a,reinecke2002b,roepke2005c,roepke2005b}.
This finite volume code is based on the \code{prometheus} \citep{fryxell1989a} implementation of the ``\textit{Piecewise Parabolic Method}'' (PPM, \citealt{colella1984a}).
It includes an appropriate equation of state for  WD matter based on the equation described by \citet{timmes2000a}. The
detonation front is modelled using the level set method \citep{reinecke1999a,
  golombek2005a, roepke2007b}, where the nuclear burning zone is
numerically treated as infinitely thin. For numerical efficiency, instead of a
full reaction network we use six
\textit{pseudo}-species, \nuc{12}{C}, \nuc{16}{O}, \nuc{20}{Ne}, \nuc{4}{He}, intermediate mass elements (IME), and iron group elements (IGE),
approximately representing fuel and ash compositions. Nuclear
statistical equilibrium is followed by a temperature and density
dependent mixture of \nuc{4}{He} and the IGE species.
The ash composition, which depends on fuel density and composition, is read off from
tables that are calibrated in a self-consistent manner as
described below (Sect. \ref{sec:calibration}).

This is sufficient to model the energetics of the detonation process. 
For convenience, the numerical simulations
presented in this work are performed in two dimensions assuming
axisymmetry. The numerical resolution is fixed to $1024 \times 1024$
grid cells that  co-expand with the explosion of the WD
so that all explosion ejecta can be followed to
homologous expansion, approximately reached $100\,
\mathrm{s}$ after ignition \citep{roepke2005c}. To determine the detailed chemical composition of the ejecta,
we apply a nucleosynthetic
postprocessing step \citep{travaglio2004a}. It is based on
approximately 90,000 tracer particles distributed in our
2D hydrodynamical simulation. This is sufficient to obtain converged nucleosynthetic yields \citep{seitenzahl2010a}.
These tracer particles are passively advected with the hydrodynamical
flow and record the thermodynamic trajectories of representative fluid
elements. In the postprocessing we follow the nuclear reactions,
using a nuclear reaction network with 384 species \citep{thielemann1986a,thielemann1990a,thielemann1996a}.

To predict observables from our explosion models we conduct radiative transfer simulations. 
Because of the spherical symmetry of the ejecta, we map the 2D distribution of the final chemical composition
and the ejecta density to a 1D grid in velocity space,
using the same smooth-particle-hydrodynamics-like algorithm
that is described in \citet{kromer2010a}. 
The radiative transfer calculations are performed with
\code{artis} \citep{sim2007b,kromer2009a}.
We use the same atomic data as in \citet{gall2012a} 
For the calculations $1.024 \times 10^7$ Monte Carlo packets are used.
The radiative transfer calculations start 2 days after explosion and end 120 days after explosion,
with the simulation discretized into 111 logarithmically separated time steps. 

\subsection{Calibration of the detonation model}
\label{sec:calibration}
\begin{figure}[]
  \begin{center}
     \includegraphics[width=0.46\textwidth]{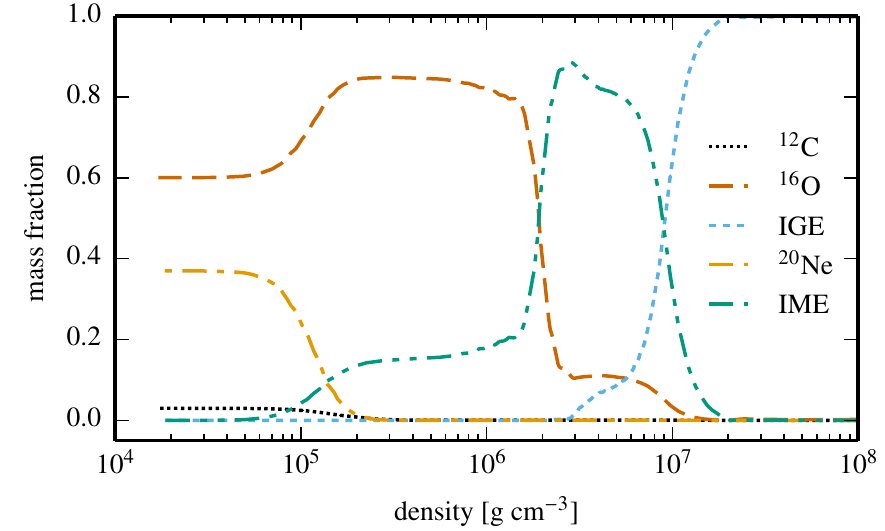}
     \caption{
       Illustration of the abundance table for the nuclear
       burning. The composition of the ash, depending on the
       density for an initial composition of 
       $X(\nuc{12}{C})=0.03$, $X(\nuc{16}{O})=0.6$, and $X(\nuc{20}{Ne})=0.37$.
     }
       \label{fig:abund_table} 
  \end{center}
\vspace{-0.0cm}
\end{figure}
The detonation in our model is not resolved but is represented by a
parameterized description. The energy release in the burning process
has to be determined and encoded in the fuel and ash composition with
the six pseudo-species. To achieve consistent energetics, we employ an
iterative calibration procedure \citep{fink2010a,ohlmann2014a}. The tracer
particles are arranged radially in the exploding star with constant distance in $\log \rho$
to ensure resolution of the transition regime between burning to
nuclear statistical equilibrium and incomplete Si burning.

In an initial run, complete burning to IGE is assumed everywhere thus
releasing the maximum amount of energy.  Subsequently, a
nucleosynthesis postprocessing step is performed that changes the ash
composition in the lower density regions. The result is mapped into
the six pseudo-species tabulated as a function of fuel density. This
serves as input for the next iteration step. The second parameter of our detonation model is the front propagation speed. It is set to the Chapman
Jouguet velocity (CJ), which is determined according to the energy
released in each iteration step.
For the calculation of the CJ speed, we use the equilibrium sound speed \citep{fickett1979a}. The procedure is repeated until the released energy in the explosion run matches 
the nucleosynthetic postprocessing result which is the
case after ten iterations.
The values of the calibrated table are plotted in Fig. \ref{fig:abund_table}.
It clearly shows different burning stages, indicated by the pseudo-species.
In the high density regime above
$\sim ~ 10^7\,\mathrm{g}\,\mathrm{cm}^{-3}$ fuel material is burned to
nuclear statistical equilibrium (NSE) resulting in IGE after
freeze-out. For intermediate fuel densities  ($10^6\lesssim \rho~
[\mathrm{g}\,\mathrm{cm}^{-3}] \lesssim 10^7$) the ash is composed of IME
and oxygen. At lower densities, carbon and neon burn to oxygen, while
below $\sim$$10^5 \, \mathrm{g}\,\mathrm{cm}^{-3}$ burning ceases.
\subsection{Simulation set-up}
\label{sec:used_models}

We calculate a series of explosion models of ONe WDs. These are set up
in hydrostatic equilibrium with central densities $\rho_0$ ranging
$\left(1 \dots 2 \right) \times 10^{8} \,\gcc$, which is well below the threshold
for electron captures to become dynamically important
(\citealt{nomoto1987a} estimates $\rho_{\mathrm{ec}} \simeq 9.5 \times
10^{9}\, \gcc$ and \citealt{canal1992a} give $\rho_{\mathrm{ec}} \simeq 8.5 \times 10^{9}\, \gcc$). 
The initial temperature of all models is assumed to be $T = 5 \times
10^{5}\, \K$ throughout the star. As in our previous work on CO WDs \citep{sim2010a}
we assume uniform composition with values motivated by the results of \citet{garcia-berro1997a} and \citet{gil-pons2001a}.
Specifically, the mass fractions of our initial composition are
$X(\nuc{12}{C})=0.03$, $X(\nuc{16}{O})=0.6$ and $X(\nuc{20}{Ne})=0.37$.
Our set-up procedure results in ONe WDs with masses of $1.18 - 1.25\,\msun$.
Their parameters are summarized in Table~\ref{tab:syntheticobservables}.
Our simulations are for zero-metallicity main-sequence progenitors. Specifically, we do not include any intial abundance of the
neutron rich isotope \nuc{22}{Ne}, which would slightly modify the results \citep{townsley2009a}.
As our intention is to study the outcome of detonations in ONe WDs
rather than their progenitor evolution and ignition, we ignite the
detonation by hand at the centre of the star. 
For comparison, we also run a simulation of a detonation in a CO WD
set up with a central density of $\rho = 1.5 \times 10^{8} \, \gcc$,
equal mass fractions of carbon and oxygen, and a uniform temperature
of $T = 5 \times 10^{5} \, \K$.

\subsection{Explosion simulations}
\label{subsec:explosion_simulations}
\begin{figure}
  \begin{center}
     \includegraphics[width=0.46\textwidth]{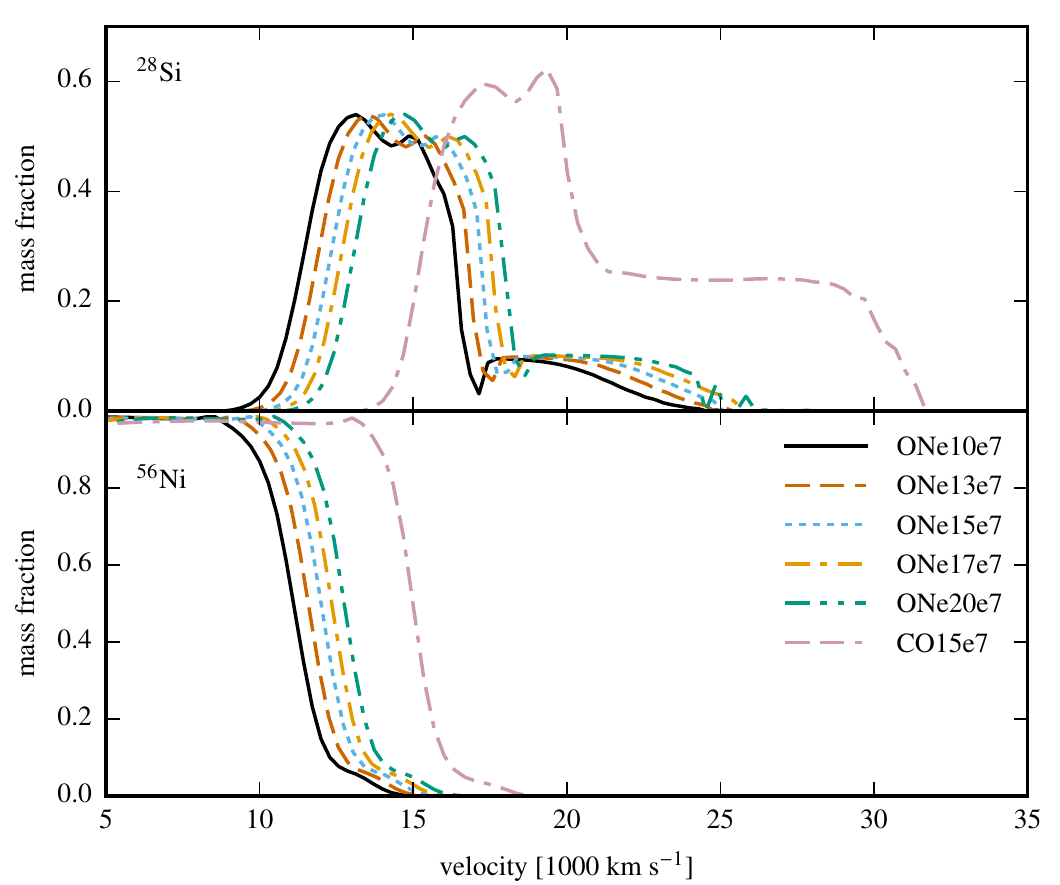}
     \caption{\nuc{28}{Si} (upper panel) and \nuc{56}{Ni} (lower
       panel) mass fractions of the ejecta in velocity space.}\label{fig:MassFraction}
  \end{center}
\vspace{-0.0cm}
\end{figure}

From our hydrodynamical simulations we find many similarities, but
also some clear differences between the CO WD and the ONe WD
detonations. Table \ref{tab:syntheticobservables} shows the results in terms of
kinetic energy of the ejecta and nuclear abundances of the pseudo-species.
As expected from the differences in the binding energy of the
fuel material relative to the ash \citep[the energy release in burning
\nuc{12}{C} to \nuc{56}{Ni} is about 30\% higher than for burning
\nuc{16}{O} or \nuc{20}{Ne} to \nuc{56}{Ni}; see][]{wang2012c} the simulations lead to a significantly lower
kinetic energy of the ejecta for ONe WD detonations. The nucleosynthetic postprocessing results (Table
\ref{tab:syntheticobservables}) show that there is a little less
$\nuc{56}{Ni}$ and overall IGE in the ejecta of the ONe WD detonation
than in the ejecta of the CO WD explosion with the same initial mass. In
contrast, the IME fraction of the ejecta is somewhat enhanced in the
ONe detonation. Both effects are, however, not very pronounced. For
the observables, the more important difference is in the distribution
of these species in velocity space. This is shown for \nuc{56}{Ni} and
\nuc{28}{Si} as representative examples in Fig.~\ref{fig:MassFraction}. The
distribution of both species is shifted towards lower velocities for
the ONe WD explosion.
Given that the \nuc{56}{Ni} masses are similar but the ejecta
velocities are lower, we expect that the models will have similar
brightness but the ONe WDs will give rise to smaller blueshifts of
spectral features. Because of the lower expansion velocities, we expect an increase in the light curve rise time.



\subsection{Comparison of synthetic spectra and observations }

\begin{figure}[]
  \begin{center}
     \includegraphics[width=0.98\columnwidth]{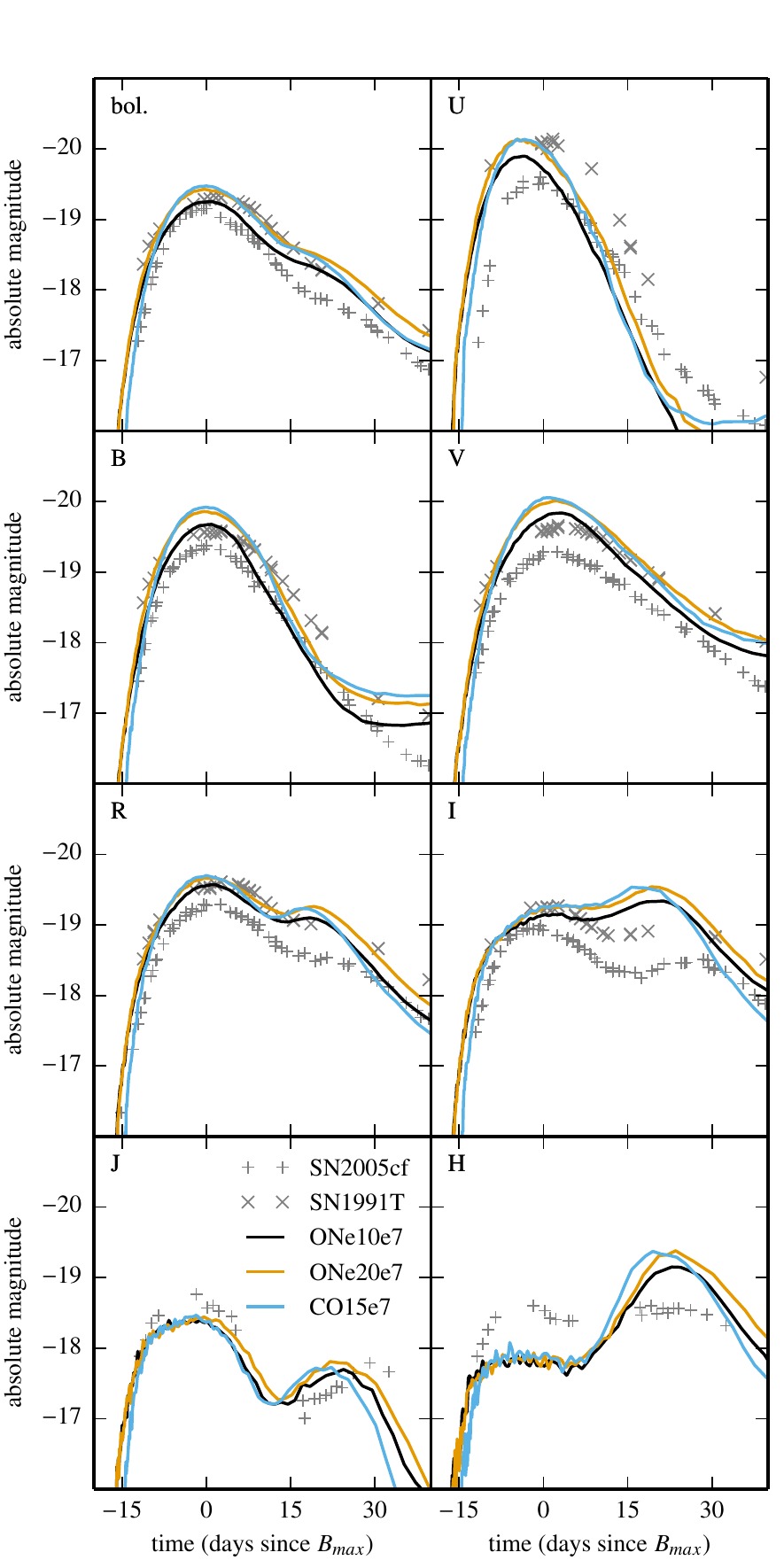}
     \caption{
       Synthetic light curves in different bands for
       our ONe models with a central density of $1 \times
       10^8~\mathrm{g~cm}^{-3}$ and $2 \times 10^8~\mathrm{g~cm}^{-3}$
       and our CO model compared to the observed light curves of
       SN~2005cf and SN~1991T.
       The SN~2005cf lightcurve is dereddened
       for $E(B-V) = 0.097$ \citep{pastorello2007a} and the SN~1991T
       lightcurve is dereddened using $E(B-V) = 0.13$ \citep{phillips1992a}.
       The distance modules are 30.76 for SN~1991T \citep{saha2006a}
       and 32.51 for SN~2006cf \citep{pastorello2007a}.
       For the bolometric lightcurve of SN~1991T we used a NIR correction based on normal SNe Ia.
     }
       \label{fig:lightcurve}
  \end{center}
\vspace{-0.2cm}
\end{figure}


\begin{table*}
  \begin{center}
  \caption{Upper half: model parameters and most abundant nuclei of our nuclear network calculations.
           The first five models are for ONe WDs with a homogeneous composition of 3\% \nuc{12}{C}, 60\% \nuc{16}{O}, and 37\% \nuc{20}{Ne} by mass.
           The last model, a homogenous CO WD with equal mass of \nuc{12}{C} and \nuc{16}{O}, is for comparison.
           Lower half: time of \band{B}-band maximum after explosion, absolute model magnitudes, $\Delta m_{15}(B)$, and
           wavelength and corresponding blueshift velocity where the absorption of the Si\,{\sc ii}\,6355\AA\ feature is deepest at \band{B}-band maximum.}
  \begin{tabular}{ccccccc}
    \hline\hline
    Model & ONe10e7 & ONe13e7 & ONe15e7 & ONe17e7 & ONe20e7 & CO15e7 \\\hline
    $\rho_{0}$ $[\gcc]$  & $1.0 \times 10^8$ & $1.3 \times 10^8$ &$1.5 \times 10^8$ &$1.7 \times 10^8$ &$2.0 \times 10^8$ &$1.5 \times 10^8$ \\
    $M_{\mathrm{tot}}$ $[\msun]$ & 1.18 & 1.21 & 1.23 & 1.24 & 1.25 & 1.23 \\\hline
    $E_{\mathrm{kin}}$ [\foe]  & $1.14 $ & $1.17 $ & $1.19 $ & $1.21 $ & $1.22 $ & $1.52 $ \\
    IME [\msun]          & 2.72e-1 & 2.28e-1 & 2.07e-1 & 1.89e-1 & 1.68e-1 & 1.66e-1 \\
    IGE [\msun]          & 8.53e-1 & 9.40e-1 & 9.82e-1 & 1.02e0  & 1.06e0  & 1.03e0  \\\hline
    \nuc{16}{O} [\msun]  & 5.20e-2 & 4.06e-2 & 3.52e-2 & 3.12e-2 & 2.65e-2 & 2.24e-2 \\
    \nuc{24}{Mg} [\msun] & 7.21e-3 & 5.72e-3 & 4.98e-3 & 4.41e-3 & 3.76e-3 & 6.10e-3 \\
    \nuc{28}{Si} [\msun] & 1.38e-1 & 1.16e-1 & 1.05e-1 & 9.57e-2 & 8.47e-2 & 9.30e-2 \\
    \nuc{32}{S} [\msun]  & 8.76e-2 & 7.37e-2 & 6.68e-2 & 6.12e-2 & 5.43e-2 & 4.80e-2 \\
    \nuc{36}{Ar} [\msun]  & 1.95e-2 & 1.65e-2 & 1.51e-2 & 1.38e-2 & 1.24e-2 & 9.91e-3 \\
    \nuc{40}{Ca} [\msun] & 1.95e-2 & 1.65e-2 & 1.51e-2 & 1.39e-2 & 1.24e-2 & 9.39e-3 \\
    \nuc{52}{Fe} [\msun] & 9.21e-3 & 8.42e-3 & 8.13e-3 & 7.95e-3 & 7.83e-3 & 4.54e-3 \\
    \nuc{56}{Ni} [\msun] & 8.32e-1 & 9.16e-1 & 9.57e-1 & 9.90e-1 & 1.03e0  & 1.00e0  \\
    \nuc{57}{Ni} [\msun] & 6.43e-3 & 7.72e-3 & 8.39e-3 & 8.98e-3 & 9.74e-3 & 1.19e-2 \\\hline
    $t_{\band{B}_{\mathrm{max}}} [\mathrm{days}]$ & 19.1 & 18.9 & 18.7 & 18.4 & 18.4 & 17.3 \\\hline
    $M_{\band{B}_{\mathrm{max}}}$ & -19.68 & -19.74 & -19.78 & -19.83 & -19.86 & -19.93 \\
    $\Delta m_{15}(\band{B})$ &  1.51 & 1.47 & 1.45 & 1.40 & 1.41 & 1.67 \\\hline
    $M_{\band{U}}(t_{\band{B}_{\mathrm{max}}})$ & -19.72 & -19.80 & -19.86 & -19.93 & -19.95 & -19.98 \\
    $M_{\band{V}}(t_{\band{B}_{\mathrm{max}}})$ & -19.79 & -19.86 & -19.91 & -19.94 & -19.98 & -20.05 \\
    $M_{\band{R}}(t_{\band{B}_{\mathrm{max}}})$ & -19.57 & -19.62 & -19.64 & -19.65 & -19.67 & -19.69 \\
    $M_{\band{I}}(t_{\band{B}_{\mathrm{max}}})$ & -19.13 & -19.17 & -19.18 & -19.20 & -19.23 & -19.25 \\
    $M_{\band{J}}(t_{\band{B}_{\mathrm{max}}})$ & -18.40 & -18.41 & -18.42 & -18.44 & -18.46 & -18.37 \\
    $M_{\band{H}}(t_{\band{B}_{\mathrm{max}}})$ & -17.76 & -17.80 & -17.81 & -17.80 & -17.81 & -17.83 \\
    $M_{\band{K}}(t_{\band{B}_{\mathrm{max}}})$ & -17.70 & -17.76 & -17.78 & -17.83 & -17.79 & -17.81 \\
    $M_{\band{Bol.}}(t_{\band{B}_{\mathrm{max}}})$ & -19.38 & -19.44 & -19.47 & -19.51 & -19.53 & -19.56 \\\hline
    $v_{\mathrm{Si~II}}$ [$10^9$ cm s$^{-1}$] & 1.3 & 1.3 & 1.4 & 1.4 & 1.4 & 1.7 \\
    $\lambda_{\mathrm{Si~II}}$ [$\AA$] & 6090 & 6080 & 6060 & 6060 & 6060 & 6000 \\\hline
  \end{tabular} 
  \label{tab:syntheticobservables}
\end{center}
\end{table*}

To compare our models with observations we have calculated synthetic
spectra and light curves. Table \ref{tab:syntheticobservables} gives
important values from these calculations: the time of maximum brightness in
\band{B} band ($t_{B_{\mathrm{max}}}$), the \band{B}-band decline time scale ($\Delta m_{15}(B)$),
and the magnitudes at $t_{B_{\mathrm{max}}}$ in the \band{U},
\band{V}, \band{R}, \band{I}, \band{J}, \band{H}, \band{K} bands and
in bolometric light. Light curves for a subset of our models are shown (and compared to
observations) in Fig. \ref{fig:lightcurve}. There is relatively little
variation among the light curves, as one would expect given that the
masses of the models are similar.  The difference in the velocity
structures of the CO WD and the ONe WD detonations lead to a rise time that is 
approximately two days longer  for the latter. The time scales
of the infrared light curve evolution are also slower in the ONe
models (e.g. the secondary IR peak is reached a little later).


Owing to their large \nuc{56}{Ni} masses, our models are too bright to
provide a good match to a normal SN~Ia such as SN~2005cf
(\citealt{pastorello2007a,garavini2007a}, Fig. \ref{fig:lightcurve}).
We do, however, find better agreement 
with the light curves of SN~1991T \citep{filippenko1992a,phillips1992a,ruiz-lapuente1992a,lira1998a}.

\begin{figure}
  \begin{center}
     \includegraphics[width=\linewidth]{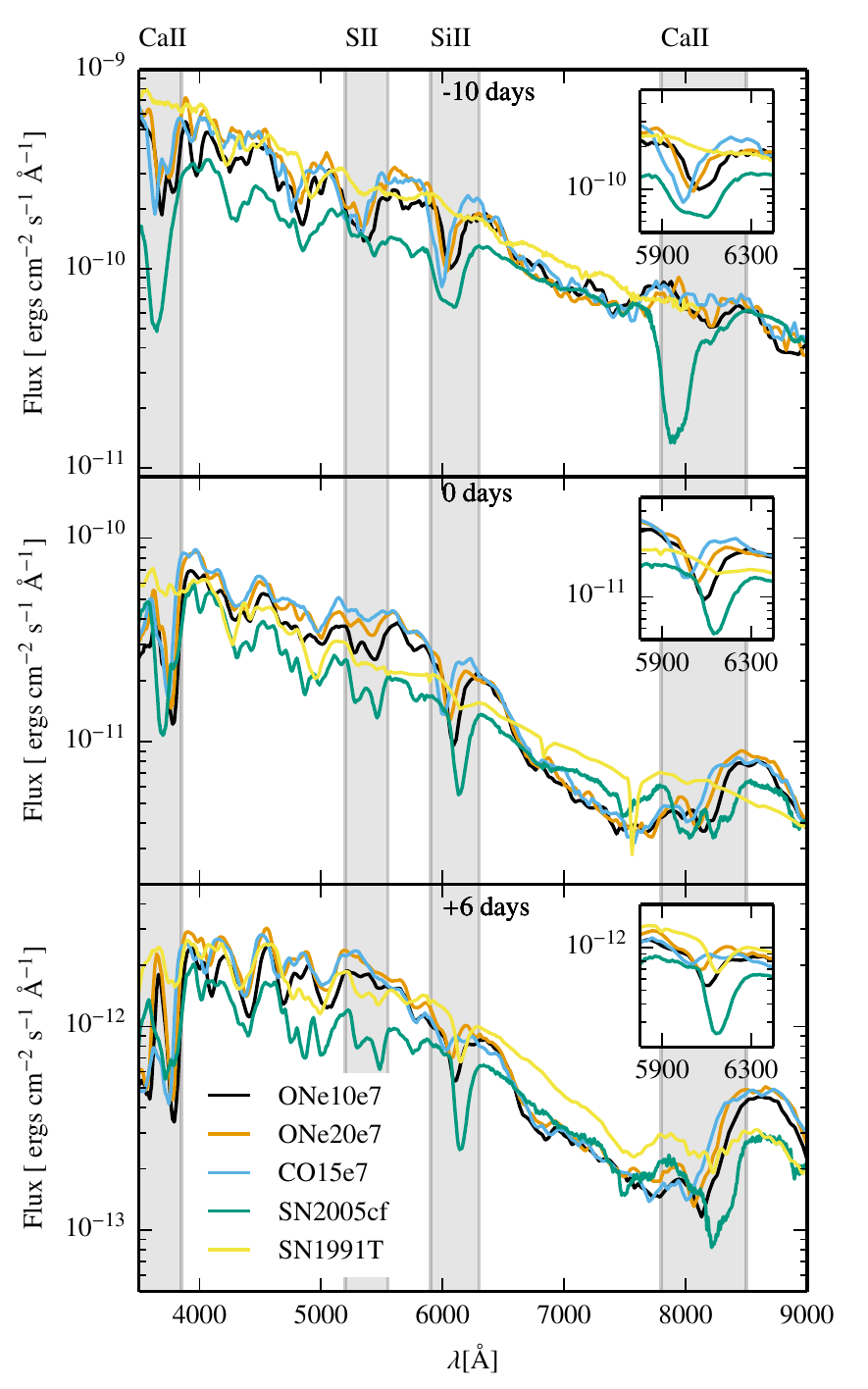}
     \caption{
              Synthetic spectra for a subset of our detonation models
              as outlined by the labels in the bottom panel.
              For comparison,  observed spectra of SN 2005cf
              (-10, 0, +6 d) and SN 1991T (-10, -3, +6 d) are shown.
              All times are relative to maximum brightness in B band.
              The spectra are de-redshifted: For SN2005cf by $z=0.00646$ and
              for SN~1991T by z=0.006059.
            }
       \label{fig:spectra}
  \end{center}
\vspace{-0.0cm}
\end{figure}

In Fig. \ref{fig:spectra}, model spectra are shown at three epochs
($-10$, 0, 6 days relative to $t_{B_{\mathrm{max}}}$).
Overall, the spectra are similar and compare equally well to the
observations; in particular, the important Si and S features are
clearly present in the models with a strength comparable to the
SN~2005cf data. There are, however, differences between the models in detail:
\begin{compactitem}
\item around maximum light the ONe WD
  detonation models show slightly stronger IME features;
\item the ONe WD explosion models have lower Si velocities (about
  $\approx 400~ \mathrm{km}\,\mathrm{s}^{-1}$; see Table
  \ref{tab:syntheticobservables}), compared to the CO WD explosion model.
  \end{compactitem}

These results (see Table \ref{tab:syntheticobservables}) confirm what we expect from the hydrodynamical simulations (as discussed above).
In all the models the blueshifts (e.g. of the Si\,{\sc ii} features) are generally too high compared to the observations, but this is less pronounced in
the ONe WD detonations (Fig. \ref{fig:spectra}).

The spectral features of our models are much too strong for SN~1991T.
This holds in particular for the early epochs ($-10$ d, 0 days after $t_{B_{\mathrm{max}}}$)
where our models show strong Si\,{\sc ii} and Ca\,{\sc ii} features, while the spectra of SN~1991T
do not show any strong absorption lines. This discrepancy is consistent with expectations based
on previous modelling of SN~1991T; in particular, \citet{sasdelli2014a} have shown that the Si in SN~1991T
should be predominantly located at velocities below $\sim 12500\,\kms$.
In contrast, our models are Si-rich out to $\sim 17000\,\kms$.
Thus, despite having appropriate brightness, the spectra show that our
models do not reproduce SN~1991T (or 91T-like objects in general).

\section{Conclusions}
\label{sec:conclusions}

Our population synthesis calculations show that the fraction of
potential SN~Ia progenitors involving ONe WDs
is non-negligible. For the double-detonation 
scenario, they account for up to 3\% of the sub-\mch\ systems (see  Table \ref{tab:rates}).
For double degenerate mergers, those involving ONe WDs contribute 4\%. If we restrict this census to systems that would reach
sufficient brightness to produce a SN~Ia
event, this fraction increases to 11\%.
In conclusion, explosions of ONe WDs are not a dominant
channel of SNe~Ia, but our results demonstrate that their contribution is important.
It is therefore worthwhile to determine possible outcomes of thermonuclear explosions in ONe WDs
in order to establish whether they can be identified as a subpopulation in transient surveys. 

We performed simulations of detonations in ONe WDs of different
masses and compared these models to detonations in a relatively massive
sub-\mch~CO WD. 
As expected, 
the kinetic energy of the ejecta of ONe WD detonations is lower than that
of an equal-mass CO WD.
There is, however, very little difference between the composition of the ejecta. 
Although overall there is no significant improvement or deterioration
in matching observed SNe~Ia, the variation in the velocity
distribution of species in the ejecta leads to changes in the
predicted observables, most importantly in the rise time of the light
curves and Si line velocities (both are slower for ONe WD than for CO
WD detonations).
Thus, with respect to spectral line shifts, our ONe models fit the
observations better than our CO models.
In terms of brightness, however, all our models are too bright compared to normal SNe~Ia.


The question of whether particular explosion triggering mechanisms can be successful in ONe WDs remains open and has to be addressed in future work.

\begin{acknowledgements}
  This work was supported by the DFG via the graduate school
  ``Theoretical Astrophysics and Particle Physics'' at the University of W\"urzburg (GRK 1147).
  Parts of this research were conducted by the Australian Research Council Centre of Excellence
  for All-sky Astrophysics (CAASTRO) through project number CE110001020 and by the the ARC Laureate Grant FL0992131.
  F.~K.~R.\ was supported by the DFG via the Emmy Noether Programme (RO 3676/1-1) and
  by the ARCHES prize of the German Federal Ministry of Education and Research (BMBF),
  S.~A.~S.\ by STFC grant ST/L000709/1,
  S.~T.~O.\ by the Studienstiftung des deutschen Volkes and
  R.~P.\ by ERC-StG grant EXAGAL-308037.
  We also thank the DAAD/Go8 German-Australian exchange programme for travel support and
  the Partner Time Allocation (Australian National University),
  the National Computational Merit Allocation and the Flagship Allocation Schemes
  of the NCI National Facility at the Australian National University. 
\end{acknowledgements}

\bibliographystyle{aa} \bibliography{astrofritz}

\end{document}